\pdfoutput=1
\documentclass[10pt,conference]{IEEEtran}
\IEEEoverridecommandlockouts
\usepackage{url}
\usepackage{cite}
\usepackage{amsmath,amssymb,amsfonts}
\usepackage{algorithmic}
\usepackage{graphicx}
\usepackage{textcomp}
\usepackage{xcolor}
\usepackage{booktabs}
\usepackage{pifont}
\usepackage{caption}
\usepackage{microtype}
\usepackage[normalem]{ulem}
\usepackage{enumitem}
\usepackage{makecell}
\usepackage{multirow}
\usepackage[hidelinks]{hyperref}
\usepackage{newtxtext,newtxmath}
\usepackage{newtxtext}
\usepackage{microtype}

\usepackage{tikz}
\usetikzlibrary{arrows.meta, positioning}
\newlength{\mainw}
\newlength{\branchw}
\newlength{\branchshift}
\usepackage{doi}
\usepackage{tcolorbox}
\tcbuselibrary{skins}

\def\BibTeX{{\rm B\kern-.05em{\sc i\kern-.025em b}\kern-.08em
    T\kern-.1667em\lower.7ex\hbox{E}\kern-.125emX}}

\newcommand{\countobservations}{
    \def \countobservations{1}
}
\newcounter{observation}
\countobservations

\newcommand{\countimplications}{
    \def \countimplications{1}
}
\newcounter{implication}
\countimplications

\newboolean{showcomments}
\setboolean{showcomments}{true}

\ifthenelse{\boolean{showcomments}}
{\newcommand{\nbc}[3]{
 {\colorbox{#3}{\bfseries\sffamily\scriptsize\textcolor{white}{#1}}}
 {\textcolor{#3}{\sf\small$\blacktriangleright$\textit{#2}$\blacktriangleleft$}}
 }
}
{\newcommand{\nbc}[3]{}
 
 }

\DeclareMathOperator{\rank}{rank}

\graphicspath{{figures/}}

\begin{document}
\title{SMTpip: Interpreter-Aware SMT-Based Dependency Conflict Resolution for Restoring Python Source-Code Executability}

\author{\IEEEauthorblockN{Sadman Jashim Sakib}
\IEEEauthorblockA{
 \textit{University of Windsor}\\
 Windsor, Canada \\
 sakib51@uwindsor.ca}
 \and
\IEEEauthorblockN{Muhammad Asaduzzaman}
\IEEEauthorblockA{
 \textit{University of Windsor}\\
 Windsor, Canada \\
Muhammad.Asaduzzaman@uwindsor.ca}
 \and
\IEEEauthorblockN{Curtis Bright}
\IEEEauthorblockA{
 \textit{University of Waterloo}\\
 Waterloo, Canada \\
 cbright@uwaterloo.ca}
}

\maketitle
\begin{abstract}
Software developers rely on packages to reuse existing functionality instead of implementing everything from scratch. Python developers commonly provide package and interpreter dependencies using configuration files, such as \textit{requirements.txt} or \textit{setup.py}. Package managers in Python, such as pip, can install packages according to dependency and interpreter version constraints specified in configuration files. However, Python dependency resolution remains challenging: (1) different packages may require incompatible versions of the same dependency; (2) dependencies may require a Python interpreter version that is incompatible with the interpreter used for the project, making a valid environment impossible; and (3) pip, the most popular Python package manager, resolves conflicts via backtracking, repeatedly trying candidate versions without knowing whether a valid execution environment exists or not. To address these challenges, we present \textbf{SMTpip}, an interpreter-aware environment inference technique for improving the executability of Python source-code artifacts. SMTpip constructs a dependency knowledge graph using metadata stored in the Python Package Index (PyPI) that hosts millions of package releases, encodes both package version constraints and interpreter compatibility constraints specified in configuration files into \textbf{Satisfiability Modulo Theories (SMT)} formulas. Solving these formulas identifies a set of package versions and an interpreter version that jointly satisfy all declared constraints. Empirical evaluation on multiple datasets from open-source Python projects shows that SMTpip achieves substantial speedups---6.9$\times$ over pip, 9.6$\times$ over Conda, 3.2$\times$ over smartPip, and 4$\times$ over PyEGo---while consistently producing constraint-consistent environments. Our execution-based evaluation shows that SMTpip enables more successful executions and substantially fewer interpreter-related failures than baseline tools. The tool and datasets are publicly available in a replication package.

\end{abstract}


\section{Introduction}
\label{sec:introduction}

Software developers rely on third-party packages to reuse functionality and accelerate development. In Python projects, dependency requirements are recorded in configuration files such as \textit{requirements.txt} and \textit{setup.py}~(Figure \ref{fig:config_files}). These files serve as a blueprint for recreating the software environment by specifying required packages and version constraints. Python packages are distributed through the Python Package Index (PyPI)~\cite{PyPI}, a central repository that hosts third-party packages. Package managers (such as \textit{pip}~\cite{pip} and \textit{Conda}~\cite{conda}) use these configuration files to download and install specified packages and any other packages that are dependent on them maintaining their version constraints. However, the installation of packages may be affected by dependency conflict issues when multiple version constraints are specified for the same package, potentially leading to build failures.

\begin{figure}[t]
\centering
\includegraphics[width=\columnwidth]{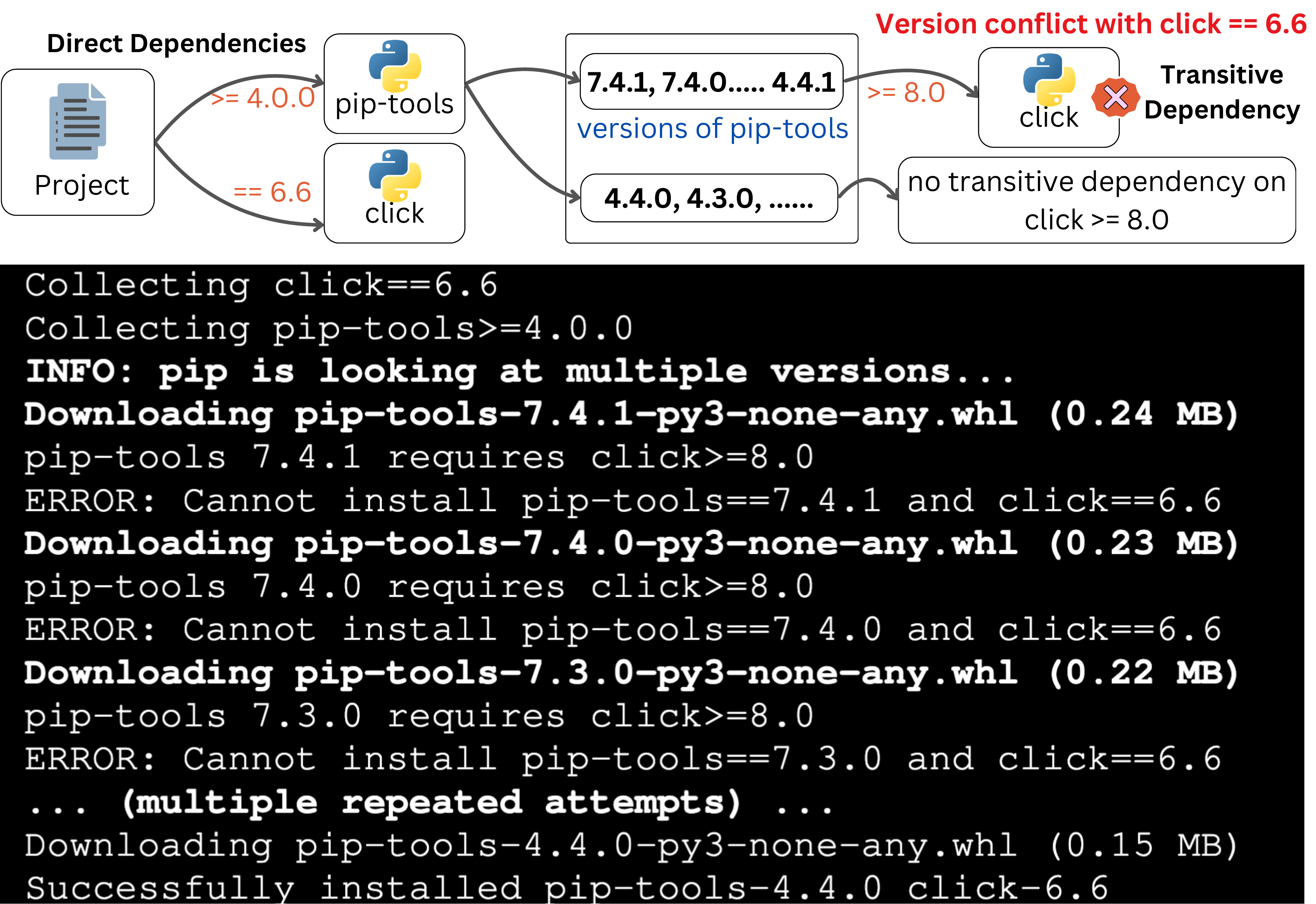}
\caption{An example of a dependency conflict and how pip resolves it using backtracking.}
\label{fig:Conflict}
\vspace{-0.5cm}
\end{figure}

For example, installing the Python project \textit{fflpy} can lead to a dependency conflict (issue~\#1~\cite{fflpy}). During installation of dependent packages exactly one version per required package must be selected such that all direct and transitive version constraints, as well as Python interpreter compatibility constraints, are satisfied. Figure~\ref{fig:Conflict} illustrates a dependency conflict in this setting: the project requires $\textit{click} == 6.6$ and also requires $\textit{pip-tools} \geq 4.0.0$, but newer \textit{pip-tools} releases (e.g., 7.4.1) introduce a transitive constraint $\textit{click}\geq8.0$, making those candidate versions of \textit{pip-tools} incompatible with the project’s pinned \textit{click} requirement.  In this scenario, pip first installs $\textit{click}$ version $6.6$ as specified in the configuration file. Next, pip attempts to install the latest available \textit{pip-tools} version (7.4.1) as it meets the version constraint ($\textit{pip-tools} \geq 4.0.0$). However, $\textit{pip-tools}$ version $7.4.1$ has a dependency on $\textit{click} \geq 8.0$. Since the project explicitly requires $\textit{click} == 6.6$, pip detects a dependency conflict and backtracks to the previous candidate version (7.4.0). Pip continues this trial-and-error process over other candidate versions until it reaches $\textit{pip-tools}==4.4.0$, which does not impose a conflicting constraint on \textit{click}, thereby resolving the dependency conflict. However, pip does not know in advance how many candidate versions it needs to try, how much computation is required, and whether a valid execution environment exists.

Existing package managers can struggle with handling dependency conflicts. For example, pip can take a significantly long time to resolve a dependency conflict as shown in the previous example. Conda is another widely used package manager (popular in scientific computing) that also performs dependency resolution. However, during our evaluation we found that Conda's classic solver configuration incurs substantial overhead in resolving dependency conflicts. We use the classic solver because it employs SAT-based dependency resolution, making it a directly relevant baseline for SMTpip's CNF-based formulation. Conda also provides other solver implementations, such as libmamba, but evaluating alternative solvers is outside the scope of this study. Conda’s developers have also acknowledged that, in practice, Conda can be slower than pip for dependency resolution~\cite{conda_issue_1,conda_issue_2,conda_issue_3,conda_issue_4}. Several prior studies also address dependency conflict issues. For example, PyEGo~\cite{Ye_KnowledgeBasedDependencyInference_2022} focuses on executability by inferring environments for Python scripts, but it does not comprehensively resolve transitive dependency conflicts at install time (and may still rely on pip's backtracking during installation). SmartPip~\cite{Wang_smartPip_2022} uses a pre-built knowledge base and \emph{satisfiability modulo theories} (SMT) solving (i.e., constraint solving over logical formulas with background theories), but its constraints are not expressed in conjunctive normal form (CNF)\@.
Because modern SMT solvers internally require CNF encodings, non-CNF encodings require a translation step that introduces overhead.
SmartPip also generally installs extra packages that are not needed.

To address these challenges and to support the executability of Python source code, we propose \textbf{SMTpip}, an SMT-based dependency conflict resolution technique that (i) builds a dependency knowledge graph from PyPI metadata, (ii) encodes dependency and interpreter constraints directly in a compact CNF form, and (iii) uses a weighted Max-SMT objective to select preferred solutions (e.g., favoring newer compatible versions and avoiding unnecessary installations). Direct CNF generation avoids relying on internal CNF transformations that introduce auxiliary variables; additional variables require the solver to perform additional propagation, the cost of which dominates SAT/SMT solving time~\cite{Prestwich_2009,een2005}. SMTpip avoids the repeated trial-and-error process inherent in pip’s backtracking-based dependency conflict resolution.
Our evaluation spans four different datasets: three of those datasets are collected from previous studies (WatchMan~\cite{watchman2020}, HG2.9K~\cite{Horton_Gistable_2018}, and SD~\cite{Ye_KnowledgeBasedDependencyInference_2022}) and a new dataset of 1,359 real-world Jupyter Notebook projects curated from GitHub. Across these datasets, SMTpip achieves substantial speedups over existing tools (6.9$\times$ over pip, 9.6$\times$ over Conda, 3.2$\times$ over smartPip, and 4$\times$ over PyEGo) in dependency resolution time (RQ2), while maintaining strong resolution coverage (RQ1). In addition, we validate whether the environments generated by SMTPip improve the executability of source code (RQ3). Results from the study show that environments produced by SMTpip enable more successful program executions and substantially fewer interpreter-related failures than baseline tools.

This paper makes the following contributions:
\begin{itemize}
  \item \textbf{SMTpip.} A Python dependency resolver that computes an \emph{executable environment} by jointly selecting compatible package versions and a Python \emph{interpreter version} that satisfy all declared constraints, or reports when no such environment exists due to conflicting constraints. It uses a direct CNF encoding with a weighted Max-SMT formulation to efficiently optimize version selection and improve executability.
    \item \textbf{Datasets.} A new benchmark of 1,359 GitHub Jupyter Notebook projects, alongside three established datasets (WatchMan, HG2.9K, and SD).
    \item \textbf{Evaluation.} A comparative evaluation against pip, Conda's classic solver, smartPip, and PyEGo, including execution-based validation of resolved environments.
 \item \textbf{Tool and artifacts.} We publicly release \textbf{SMTpip}, along with all datasets and evaluation artifacts required to reproduce our results. The tool and replication package are available on Zenodo~\cite{SMTpip,SMTpipevaluation}.
\end{itemize}

The remainder of this paper is organized as follows. Section~\ref{PRELIMINARIES} introduces background and formal definitions. Section~\ref{RELATED WORK} positions SMTpip relative to prior work. Section~\ref{approach} presents our approach and encoding. Section~\ref{evaluation} reports experimental results for RQ1--RQ3. Section~\ref{sec:discussion} discusses questions related to our study and Section~\ref{THREATS} discusses threats to validity. Finally, Section~\ref{conclusion} concludes the paper.

\section{Preliminaries}
\label{PRELIMINARIES}
This section provides necessary background related to our study.

\subsection{Packages, Versions, and Constraints}
\label{sec:pkg_versions_constraints}

For each package $p$, let~$V_p$ denote the finite set of available version numbers of package~$p$.

\paragraph{Version constraints}
Say a package has a dependency on package $q$.  A \emph{version constraint} $C_q$ on $q$ defines a subset $S(C_q) \subseteq V_q$ specifying the versions of $q$ meeting the dependency. In practice, constraints are derived from PEP~440 version specifiers such as inclusive bounds ($\leq$, $\geq$), exclusive bounds ($<$, $>$), and the compatible release operator $\sim=$.

\paragraph{Third-party package dependency constraints (direct and transitive)}
For a package version $p_v$ ($v \in V_p$), let $Q$ be the set of packages $p_v$ depends on, and $C_q$ ($q\in Q$) the version constraint that $p_v$ has on package $q$.  Then
$\{\,S(C_q):q\in Q\,\}$
denotes its dependency specification, where each subset $S(C_q)$ denotes that selecting $p_v$ requires a package $q_{v'}$ to be selected where $v'\in S(C_q)$ for each $q\in Q$. Dependencies are \emph{transitive} when they are induced indirectly via other dependencies rather than listed directly by the client project.

\paragraph{Interpreter constraints}
Let $\mathcal{I}$ denote the set of available Python interpreter versions. Each package version $p_v$ may additionally define an interpreter compatibility constraint
specifying the subset of Python interpreter versions $\psi(p_v)\subseteq\mathcal{I}$ that can be used with package~$p_v$.

\subsection{Dependency Conflict}
\label{sec:dependency_conflict}

\paragraph{Third-party package conflict}

A third-party package conflict may occur in the process of selecting a set of package versions meeting all dependency requirements simultaneously.  A \emph{dependency conflict} happens when a set of selected package versions requires a package~$q$ to be installed, but there is no consistent way of selecting a package version of $q$ meeting all dependency constraints simultaneously.

For example, in Figure~\ref{fig:Conflict}, the project has constraints $\textit{click} ==6.6$ and $\textit{pip-tools} \geq 4.0.0$. Candidate versions of \textit{pip-tools} such as 7.4.1 impose a transitive constraint $\textit{click}\geq8.0$, making those candidates conflicting with $\textit{click} == 6.6$. However, older candidates (e.g., $\textit{pip-tools} == 4.4.0$) remain compatible, so the overall instance is satisfiable even though several candidate versions of {pip-tools} will result in a dependency conflict if they are installed.

\begin{figure}
\centering
 \includegraphics[width=1\linewidth]{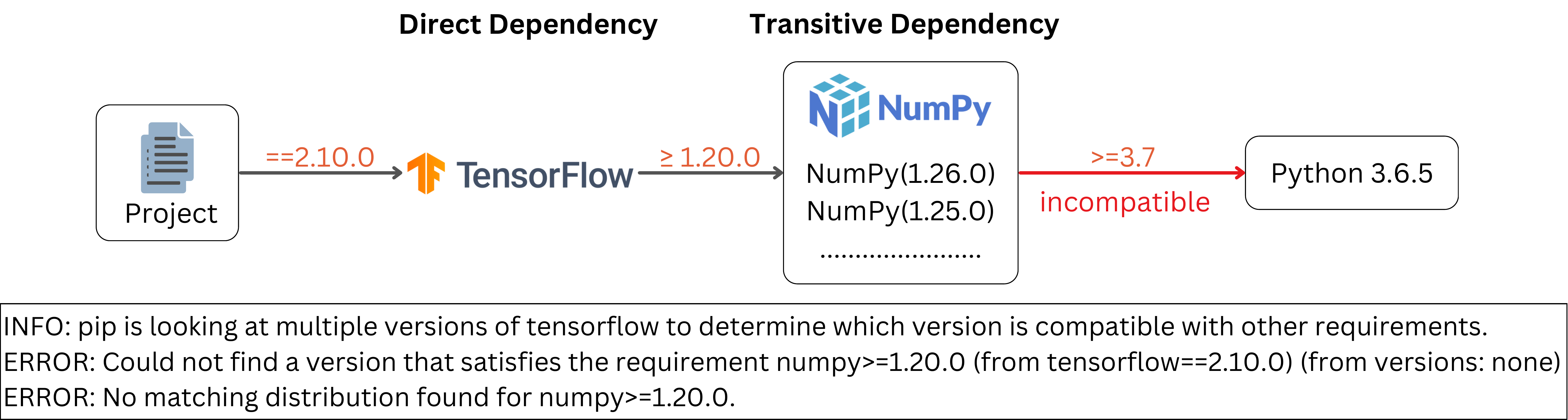}
 \caption{An example of installation failure due to Python version incompatibilities.}
 \label{Fig:incompitable}
 \end{figure}

\paragraph{Interpreter compatibility conflict}
An \emph{interpreter compatibility conflict} occurs when an interpreter constraint causes there to be no consistent way of installing packages meeting the constraint. For example, in Figure~\ref{Fig:incompitable} a project requires $\textit{TensorFlow}==2.10.0$, which depends on $\textit{numpy}\geq 1.20.0$. However, all versions of $\textit{numpy}\geq 1.20.0$ require $\textit{Python}\geq 3.7.0$. Therefore, if the environment uses Python 3.6.5, no version of NumPy can satisfy both the dependency and interpreter constraints, and the installation fails.

\subsection{Dependency Resolution and Environment Setup}
\label{sec:resolution_envsetup}

\emph{Dependency resolution} selects exactly one version for each required package, and one Python interpreter version $I$, so that: (i) all direct constraints are satisfied, (ii) all transitive dependency constraints induced by selected versions are satisfied, and (iii) all interpreter constraints hold, i.e., $I\in\psi(p_v)$ for all selected package versions $p_v$. When flexible version specifiers yield many package version candidates, dependency resolution solvers must identify and eliminate conflicting package versions until they find a constraint-consistent environment (or determine that no such environment exists).

\begin{figure}
    \centering
    \includegraphics[width=1.0\linewidth]{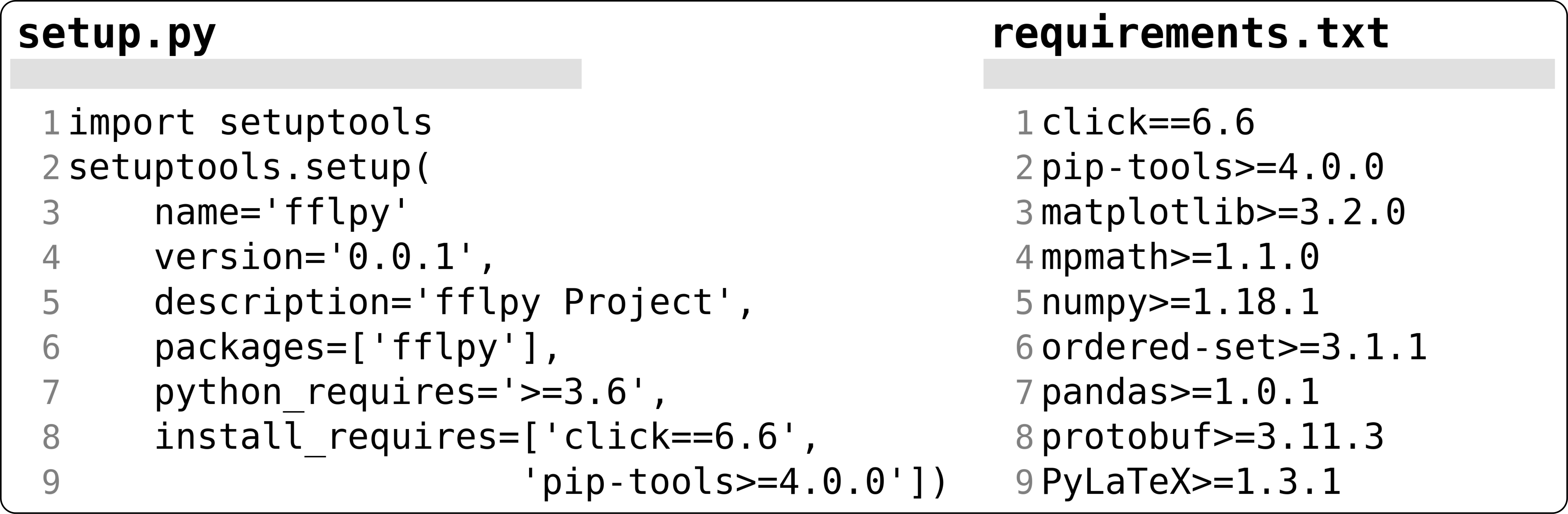}
    \caption{Examples of configuration files.}
    \label{fig:config_files}
\end{figure}

\subsection{Satisfiability Solving}
\label{sec:sat}
Satisfiability solving asks a simple question: \emph{can we assign true/false values to variables so that a logical formula becomes true?} A standard normal form for logical formulae is called \textbf{Conjunctive Normal Form (CNF)}\@. A CNF formula is a conjunction (AND, $\land$) of clauses, where each clause is a disjunction (OR, $\lor$) of literals, and a literal is either a variable or its negation.  For example, $(x\lor\lnot y)\land(\lnot x\lor z)$ is in CNF\@.
CNF is the standard input format for modern SAT solvers, as modern SAT solving algorithms operate directly on clauses.  Non-CNF encodings can be converted into CNF
by introducing new variables~\cite{Bright2020}, but this increases the formula size, and in general smaller
CNF encodings with fewer variables tend to give better solver performance~\cite{een2005}.
The notation $x\rightarrow y$ is shorthand for $\lnot x\lor y$.

\section{Related Work}
\label{RELATED WORK}

Prior work on Python environment setup and dependency management spans several research directions. We group the related works into three different categories: Dependency Resolving, Build-failure Repair, and SAT/SMT in Software Engineering.\\
\textbf{Dependency Resolving.}
A prominent line of work builds a dependency knowledge base and applies reasoning to select compatible package versions. \textbf{DockerizeMe}~\cite{DockerizeMe_2019} combines static analysis of code snippets with PyPI metadata and heuristics to assemble runnable environments, but its inferred dependency graph can be incomplete. \textbf{SmartPip}~\cite{Wang_smartPip_2022} strengthens this direction by constructing a PyPI-based knowledge graph and using an SMT solver to search for compatible versions from configuration files; however, smartPip uses a non-CNF encoding and does not explicitly compute a globally compatible Python interpreter version, which can contribute to interpreter-related failures. \textbf{PyEGo}~\cite{Ye_KnowledgeBasedDependencyInference_2022} uses an SMT solver together with code-based inference to recover environments for single-file scripts, but it is not designed for multi-file projects or configuration-driven dependency resolution; furthermore, it may leave transitive dependency conflict resolution to {pip} during installation. Other techniques such as \textbf{PyCRE}~\cite{PyCRE} and \textbf{PCREQ}~\cite{PCREQ} focus on extracting or reconstructing requirement sets, but they do not directly target efficient conflict resolution under Python’s single-version semantics. More broadly, Pinckney et al.\ formulate dependency management using SMT/Max-SMT and graph-based reasoning in \textbf{PACSOLVE}~\cite{MaxSMT_2023}; however, this work primarily targets ecosystems such as NPM where multiple versions of a package can be co-installed. This avoids many single-version conflicts, but introduces a different dependency-graph selection problem. Among these approaches, \textbf{smartPip} and \textbf{PyEGo} are the most relevant to our work because they also use solver-based reasoning to select dependency versions.

\textbf{Build-failure Repair.}
Another set of approaches attempts installation and uses build/installer feedback to propose fixes. \textbf{ReadPyE}~\cite{ReadPyE} extends PyEGo with log-driven validation to iteratively repair failing setups. \textbf{PyDFix}~\cite{PyDFix} and pip-error mining approaches (e.g., Cao et al.~\cite{help}) parse installer outputs to identify missing modules and propose dependency patches, while \textbf{V2}~\cite{V2_2019} combines a failure database with guided search over candidate versions. These methods can be effective in practice, but they often incur substantial overhead due to repeated failed installs and retries, and may struggle with nested transitive conflicts or ambiguous error messages. Complementary work addresses failures that occur after installation, such as API deprecations or missing runtime resources. \textbf{SnifferDog}~\cite{Snifferdog} targets notebooks by inferring dependencies from cells and traces, \textbf{RELANCER}~\cite{Zhu_RestoringExecutabilityJupyterNotebooks_2021} focuses on notebook API deprecations, and \textbf{LooCo}~\cite{LooCo_2023} explores constraint relaxation guided by code evidence. Outside Python, systems such as \textbf{Decca}~\cite{do} and \textbf{LibHarmo}~\cite{Huang_InteractiveLibraryVersionHarmonization_2020} harmonize versions using API-compatibility signals and cross-project usage, and SAT-based harmonization techniques have been explored (e.g.,~\cite{DHard}). Recent work also explores large language models: \textbf{PLLM}~\cite{raider} combines retrieval-augmented generation with an error-driven loop to suggest fixes, but such approaches are not fully deterministic and may hallucinate or suggest incorrect updates without structured validation. Since build failures can be caused by dependency conflicts, techniques that focus on build-failure repair can benefit from SMTpip by using its constraint-consistent dependency resolution to avoid or reduce trial-and-error installation attempts.

\textbf{SAT/SMT in Software Engineering.}
SAT/SMT solvers have also been widely applied beyond dependency management, including program analysis and verification, test generation, fault localization, and security analysis (e.g.,~\cite{Xie2005Saturn,Bjorner2011ProgrammingZ3,Si2017MaxSAT,Vanegue2012SMTSecurity}). However, the objectives of these techniques are different from ours.

\section{SMTpip: SMT-driven approach}
\label{approach}

 \begin{figure*}
\centering
 \includegraphics[scale=0.107]{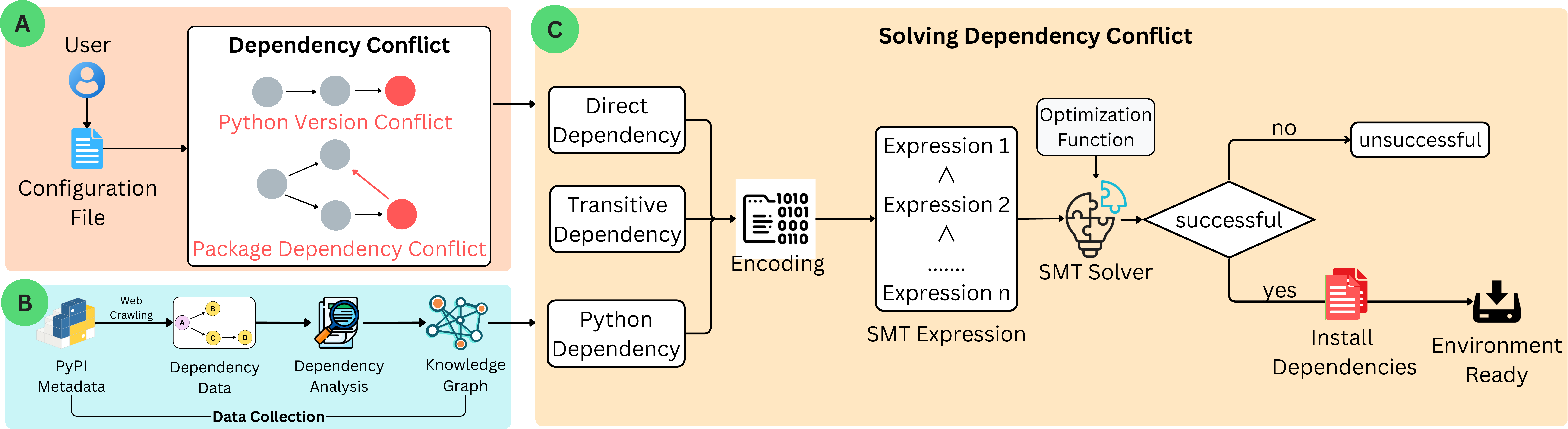}
 \caption{SMTpip architecture}
 \label{fig:approach}
 \vspace{-12pt}
 \end{figure*}
Figure~\ref{fig:approach} shows the overview of our proposed approach, SMTpip. The approach consists of two parts: knowledge graph construction and dependency resolution. This section focuses primarily on constructing the dependency knowledge graph by extracting Python version constraints and dependency version constraints from a central repository of package information. Additionally, it details the encoding of these dependency constraints into an SMT encoding.

\subsection{Knowledge Graph Construction}
\label{sec:kgraph}
Since pip with the backtracking strategy is required to iteratively download one candidate version of a concerned library when a dependency conflict occurs, it becomes inefficient when the number and size of iteratively downloaded libraries are large.
This is why we construct a knowledge graph in advance. This section describes how the dependency data is collected and modeled into a knowledge graph.

\paragraph{Data collection} Figure~\ref{fig:approach} illustrates the data collection procedure. PyPI, a central repository for third-party libraries in the Python language, is extensively used in most Python project developments. To collect the necessary data, we retrieved the dependency data from PyPI by collecting the configuration files of each library. At the time of data collection, there were around 6 million releases and nearly half a million packages available on PyPI\@. The entire process of downloading this information took 10 days.

\paragraph{Dependency analysis} Once all Python libraries were collected from PyPI, we proceeded to analyze the dependency configuration files for each package. This step involved extracting the version constraints for the dependency libraries as well as the compatible Python version. 

\paragraph{Knowledge graph representation of package dependencies} The dependency constraints were modeled into a knowledge graph representing relationships between packages, their versions, and respective dependencies. The graph comprises two types of nodes: \textbf{package nodes}, representing individual packages, and \textbf{version nodes}, representing specific versions of a package. Each version node includes two properties---\textit{python{\textunderscore}version}, denoting the required Python version, and \textit{release{\textunderscore}date}, indicating the version’s publication date on PyPI. The graph also includes two types of edges: \textbf{versioning edges}, which link package nodes to their version nodes, and \textbf{dependency edges}, which connect version nodes of dependent packages to capture inter-version dependencies. Collectively, these components form a comprehensive structure modelling the hierarchical relationships and dependencies within the Python package ecosystem. Stored in JSON format (122~MB), the knowledge graph supports efficient analysis and resolution of third-party package dependency conflicts and Python version incompatibilities.

\paragraph{Knowledge graph update}
Downloading the entire PyPI repository from scratch for every update is neither practical nor scalable due to the continuous growth and frequent changes in PyPI. To support efficient maintenance, SMTpip implements an incremental update mechanism that leverages the \texttt{last\_serial} value provided by PyPI~\cite{SMTpipKgraph}. Instead of re-fetching metadata for all packages, the updater identifies packages whose metadata has changed since the last update and retrieves only the modified records. Using this approach, a complete update scan over the entire PyPI ecosystem requires approximately 3.2 hours. Furthermore, SMTpip provides a project-level refresh mechanism through the command \texttt{python SMTpip.py -d project\_directory --refresh}, which checks PyPI only for the packages used within the target project and updates their local metadata before dependency resolution. This operation typically completes in less than one second and propagates any retrieved updates to the shared knowledge graph. For reproducibility, the generated graph is published as a JSON artifact in a GitHub repository; however, production deployments can utilize more scalable storage backends such as indexed databases or object stores. The knowledge graph could also be hosted in a server, instead of across users, avoiding the need for each developer to maintain a separate 122 MB copy.

\subsection{SMT Encoding of Dependency Resolution}
\label{sec:solving}

Solving the dependency constraints involves addressing two primary types of issues: resolving Python version incompatibilities and resolving library dependency conflicts. These two types of dependency conflicts are the focus of our work, as discussed in Section~\ref{sec:introduction}\@. In this section, we formalize the process of encoding the dependency resolution problem as a satisfiability problem. 

Given a set of dependency constraints (as shown in Figure~\ref{fig:config_files}),
our goal is to encode the dependency resolution problem as an instance of the SAT modulo theories (SMT) problem.
That is, the resulting SMT instance should have a solution exactly when the dependency resolution problem
has a solution.
A solution to the SMT instance can also be converted into a solution to the dependency resolution problem.

To do this, we define the literal $p_v$ representing that version~$v$ of package~$p$ is to be installed.
Say $V(p)$ denotes the set of all version numbers of package $p$.
Since at most one version of any package can be installed at once, we encode this constraint
using the expression $\sum_{v\in V(p)}p_v \leq 1 $.
Such expressions are natively represented by a cardinality constraint in our SMT instance.  For example, the SMT solver Z3~\cite{smt} supports the parameterized theory function ``\verb|(_ at-most 1)|'' as part of an extension of the SMT-LIB~\cite{smtlib}.

Secondly, we need to express the requirements found in the configuration file (e.g., \emph{requirements.txt}).
Each package version constraint in the configuration file determines a subset of the possible versions of package $p$ meeting that requirement.  For example, if $V(p)=\{1,2,3\}$ the requirement $\textit{p} \geq 2$ would only be satisfied
by version 2 or 3 of package $p$.
Suppose $C_p$ is a constraint in the configuration file involving package~$p$ and $S(C_p)\subseteq V(p)$ denotes
the versions of $p$ meeting the constraint.  Then we enforce that an acceptable version of $p$ will be installed
with the clause $\bigvee_{v\in S(C_p)} p_v$.
Note that the equality constraint $\textit{p} == 1$ then results in a \emph{unit clause} (a clause of length 1), i.e., just
the literal $p_1$.

Furthermore, installing a package may require other packages to be installed---these are known as \emph{transitive dependencies} and we need to ensure that all transitive dependencies are met on all packages that are installed.
Say that the configuration file of version $v$ of package $p$ requires the constraint $C_q$ on package $q$.
That is, if package $p$ version $v$ is installed, then a version package of $q$ whose version number
is in the set $S(C_q)$ must also be installed.  We represent package dependencies using the clause $p_v \rightarrow \bigvee_{u \in S(C_q)} q_u$
for each package $p$ in the original set of requirements, or in the requirements of any package that $p$ depends on, and so on inductively, until all packages the starting set may depend on are identified. These packages are determined by computing the components of the dependency knowledge graph containing the vertices corresponding to the original requirements. The graph components are computed using a recursive algorithm with a dictionary to track visited nodes. Starting from the direct dependencies, the algorithm recursively visits all connected transitive dependencies, adding each node to the dictionary once explored. If a node is encountered that is already in the dictionary, the algorithm skips further exploration of that node, ensuring efficient identification of all relevant packages without redundant visits.
The Python version constraints are encoded similarly.

Finally, we would like the solver to prioritize the most recent versions of packages (when possible) and to not install any unnecessary packages.
In order to do this, we define an objective function that the solver maximizes. This function assigns higher values to solutions selecting more recent versions and encourages not installing packages at all.
Let $P$ denote the set of all packages in the instance.
Precisely, the objective function is given by $\sum_{p \in P} \left( \sum_{v \in V(p)} \frac{\rank(p_v)}{|V(p)|} \cdot p_v + p_\text{none} \right) .$
In this expression, note that
the Boolean $p_v$ takes the value 1 if version $v$ of package $p$ is installed (and 0 otherwise),
while the Boolean $p_\text{none}$ takes the value 1 if no version of package $p$ is installed (and 0 otherwise).
Also, $\rank(p_v)$ is the rank of version $v$ of package $p$, defined such that the oldest version has rank 0, the second oldest has rank~1, and so forth, with the most recent version having rank $|V(p)|-1$.

To ensure consistency, we enforce that at most one of \(\{\,p_v \mid v \in V(p)\cup\{\text{none}\}\,\}\) is true.
This is encoded in the SMT solver using a cardinality constraint and logical implications.
Specifically, we use Z3's \verb|(_ at-most 1)| constraint to ensure that the sum of the Boolean variables \(\{\,p_v \mid v \in V(p)\,\} \cup \{p_\text{none}\}\) is at most 1, i.e., the cardinality constraint from before is modified to be \(\sum_{v \in V(p)} p_v + p_\text{none} \leq 1\). Additionally, we add the implications \(p_\text{none} \rightarrow \neg p_v\) for each version \(v\in V(p)\).
The weights \(\frac{\rank(p_v)}{|V(p)|}\) are normalized, ranging from 0 to \(\frac{|V(p)|-1}{|V(p)|} < 1\), ensuring that selecting \(p_\text{none} = 1\) gives the highest possible value of 1 in the
parenthesized expression in the objective function.
This structure incentivizes the solver to avoid installing a package unless required by the constraints, while prioritizing the most recent versions when installation is necessary.

\subsection{Example of SMT Encoding of Dependency Resolution}
Consider the open-source project fflpy~\cite{fflpy}, with the dependencies
\(\textit{click} == 6.6\) and \(\textit{pip-tools} \geq 4.0.0\).
We encode these dependencies along with all transitive dependencies and Python version requirements using Boolean variables and clauses.
Let \(c_v\) be a Boolean variable for version~\(v\) of \(\textit{click}\) (with $c_\textit{none}$ denoting no version of \(\textit{click}\) will be installed),
and \(p_v\) be a Boolean variable for version~\(v\) of \(\textit{pip-tools}\) (with $v_\text{none}$ denoting no version of \(\textit{pip-tools}\) will be installed).
Let \(A\) be the set of all \(\textit{click}\) versions (and the symbol $\text{none}$),
    \(B\) be the set of all \(\textit{pip-tools}\) versions (and the symbol $\text{none}$), and
    \(T \subseteq B\) be the versions of \(\textit{pip-tools} \geq 4.0.0\), e.g., \(\{7.4.1, 7.4.0, \dots, 4.0.0\}\).

\paragraph{Package Requirement Constraints}

\(\textit{click} == 6.6\) is encoded as $c_{6.6}$, and
\(\textit{pip-tools} \geq 4.0.0\) is encoded as $ \bigvee_{v \in T} p_v$ (e.g., $p_{7.4.1} \lor p_{7.4.0} \lor \cdots \lor p_{4.0.0}$).

 \paragraph{Consistency Constraints}
        The constraint at most one version per package is selected is encoded as $        \sum_{v \in A} c_v \leq 1 \land \sum_{v \in B} p_v \leq 1$.
        We also add the constraints
        $c_\text{none}\rightarrow \lnot c_v$ for all $v\in A\setminus\{\text{none}\}$
        and $p_\text{none}\rightarrow\lnot p_v$ for all $v\in B\setminus\{\text{none}\}$.

 \paragraph{Transitive Dependency Constraints}
    \begin{itemize}
        \item \(\textit{pip-tools} == 7.4.1\) requires \(\textit{click} \geq 8.0\) encoded as
    
        $p_{7.4.1} \rightarrow (c_{8.0} \lor c_{8.1})$
     
        \item \(\textit{pip-tools} == 4.4.1\) requires \(\textit{click} \geq 7.0\) encoded as
        
       $ p_{4.4.1} \rightarrow (c_{8.0} \lor c_{8.1} \lor c_{7.0} \lor c_{7.1})$
        
    \end{itemize}

 \paragraph{Python Version Constraints}
    \begin{itemize}
        \item \(\textit{click} == 6.6\) requires \(\textit{python} \geq 2.7\) encoded as
        
       $ c_{6.6} \rightarrow (\textit{python}_{2.7} \lor \textit{python}_{2.8} \lor \cdots)$
        
        \item \(\textit{pip-tools} == 7.4.1\) requires \(\textit{python} \geq 3.9\) encoded as
        
       $ p_{7.4.1} \rightarrow (\textit{python}_{3.9} \lor \textit{python}_{3.10} \lor \cdots)$
        
    \end{itemize}

 \paragraph{Optimization}

To encode the objective function in SMT, we formulate a MaxSMT problem using the following weighted soft constraints.
A constraint is soft if it is not strictly required, but the solver tries to satisfy as many soft
constraints as possible.
\begin{itemize}
    \item A soft clause asserting \( p_\text{none} \) with weight 1. This corresponds to the case where no version of \( \textit{pip-tools} \) is installed.
    \item For each version \( v \in T \), a soft clause asserting \( p_v \) with weight \( \frac{\rank(p_v)}{|T|} \), where \( \rank(p_v) \) is the rank of version \( v \) of \( \textit{pip-tools} \) (0 for the oldest version, 1 for the second oldest, $\dotsc$, \( |T|-1 \) for the newest).

\end{itemize}

For example, there are 47 versions of pip-tools in $T$, from 4.0.0 ($\rank(p_{4.0.0}) = 0$) to 7.4.1 ($\rank(p_{7.4.1}) = 46$). The soft constraint for $p_\text{none}$ has weight 1, while version-specific clauses have weights such as $\frac{46}{47} \approx 0.978$ for version 7.4.1 and $\frac{0}{47} = 0$ for version 4.0.0.

The solver maximizes the weighted sum of the satisfied soft constraints. Selecting $p_\text{none} = 1$ contributes a weight of~1, while installing a version $p_v = 1$ contributes a weight of $\frac{\rank(p_v)}{47} < 1$, so the soft constraints prioritize not installing a package when possible. When installation is required, newer versions with higher ranks maximize the objective function.

This encoding ensures compatibility and resolves conflicts efficiently, suitable for processing by an SMT solver.
Moreover, the dependency constraints are expressed directly in CNF\@. The consistency constraints
are expressed using cardinality constraints, but SMT solvers are designed to automatically
convert cardinality constraints into CNF efficiently.

\paragraph{Solving and Validation} Once all dependency constraints are encoded into SMT expressions, we utilize the Z3 SMT solver~\cite{smt} to determine a satisfying assignment. This assignment specifies compatible versions for the required libraries and a Python version that adheres to all constraints, ensuring a conflict-free installation.

After obtaining the solution (satisfying assignment), a validation step is performed using the knowledge graph.
In this step, all dependencies of the selected versions of the packages in the solution set are cross-verified
to ensure the packages selected to install do indeed satisfy all requirements and do not introduce any conflicts when installed together.
This validation check is not strictly required, but acts as a sanity check that the correct SMT encoding was provided to the SMT solver.
In cases where no satisfying assignment exists, the solver produces a proof of the inconsistency of the instance.

\section{Evaluation}
\label{evaluation}
This section discusses the evaluation procedure and results of our study. 

\subsection{Datasets} For the purpose of this study, we consider four different datasets (see Table~\ref{tab:dataset}). Three of those datasets (Watchman~\cite{watchman2020}, HG2.9K~\cite{Horton_Gistable_2018}, and SD~\cite{Ye_KnowledgeBasedDependencyInference_2022}) were collected from prior studies~\cite{Wang_smartPip_2022,Ye_KnowledgeBasedDependencyInference_2022,ReadPyE,DockerizeMe_2019}. We created a fourth dataset, consisting of 1,359 real-world Jupyter Notebook projects collected from GitHub, specifically curated to capture a diverse range of dependency structures found in real-world scenarios.

The Notebook dataset was derived from the GHTorent archive~\cite{GHTorent} and initially contained 247,356 Python-based Jupyter Notebook projects. We queried the GitHub API to identify projects containing \textit{requirements.txt} files, a standard indicator of dependency specifications. This filtering step reduced the dataset to 22,340 projects with dependency metadata. We then used pip to install dependencies for each project and recorded the installation logs. We analyzed the logs to detect instances in which pip invoked backtracking during dependency resolution, which we used as an indicator of dependency conflicts. The resulting Notebook Dataset comprises 1,359 real-world Jupyter Notebook projects with dependency conflicts. In total, the dataset contains 3,081 \textit{.ipynb} files.

We also measured the size of the dependency graph for each case across all datasets. The size of a dependency graph is defined as the total number of unique packages that the project depends on, directly or indirectly. For example, consider a project that depends on \textit{dagit} version 1.1.5 which in turn depends on \textit{dagster} version 1.1.5, and \textit{dagster} depends on \textit{universal-pathlib} (any version).  If there are no other dependencies, the dependency graph size will be 3. The total number of unique packages in this graph represents its size.
The maximum dependency graph sizes observed in our datasets are 85, 99, 73, and 161 for the Watchman, HG2.9K, SD, and Notebook datasets, respectively. 

\begin{table}
\caption{Datasets used for evaluation and information about the instances in each dataset.}\label{tab:dataset}
\resizebox{8cm}{!}{
\begin{tabular}{|c|c|c|}
\hline
\textbf{Dataset} &
  \textbf{Total Cases} &
  \textbf{\begin{tabular}[c]{@{}c@{}}Dependency Graph Size\\ (Min/Max/Median/Average)\end{tabular}} \\ \hline
\textit{WatchMan} & 159  & 8 / 85 / 30 / 13   \\ \hline
\textit{HG2.9K}   & 2891 & 1 / 99 / 17 / 6    \\ \hline
\textit{SD}       & 100  & 1 / 85 / 11 / 4    \\ \hline
\textit{\begin{tabular}[c]{@{}c@{}}Notebook\end{tabular}} & 1359 & 2 / 161 / 10 / 20 \\ \hline
\end{tabular}
}
\end{table}

\subsection{RQ1: How effective is SMTpip in resolving third-party package dependency conflicts and Python version incompatibilities?}
\label{sec:rq1}

 This section evaluates the performance of \textbf{SMTpip} in resolving third-party package dependency conflicts and Python version incompatibilities during installation, compared to four existing tools: pip, Conda, smartPip, and PyEGo. Results are presented in two parts: the first part assesses the tools’ ability to resolve dependency conflicts using the \textit{latest dependency knowledge graph}, as shown in Table~\ref{tab:latest_combined} to compare with pip, Conda, and PyEGo. The replication package of smartPip relies on a dependency knowledge base that was last updated in 2022. Since dependency resolution may be affected by the availability of newer dependency metadata, using more recent information could introduce bias in the comparison. To ensure a fair comparison with smartPip, we restrict the dependency information used in our experiments to versions available up to 2022, aligning with the temporal scope of SMTpip’s knowledge base. Thus, the second part discusses the evaluation conducted using a \textit{downgraded knowledge graph} to compare with smartPip, as shown in Table~\ref{tab:downgraded_combined}. In the tables, the column \emph{Total Time (sec)} represents the time required by each tool to perform dependency resolution for all projects in each benchmark. This measurement excludes time spent downloading packages or performing installations, as those operations are not considered part of the resolution process itself.

SMTpip demonstrates superior performance in resolving dependency conflicts, as evidenced in Table~\ref{tab:latest_combined} and Table~\ref{tab:downgraded_combined}. Using the latest knowledge graph, SMTpip achieves resolution rates of \textbf{72\%, 58\%, 76\%, and 66\%} across WatchMan, HG2.9K, SD, and Notebook datasets, respectively. For example, in the HG2.9K dataset, SMTpip resolves 1,668 cases in \textbf{433.68 seconds}. In contrast, pip, while achieving 58\% resolution for HG2.9K (1,668/2,891), requires \textbf{3,185.20 seconds} due to exhaustive backtracking and repeated metadata downloads. Conda resolves only 37\% of HG2.9K cases (1,062/2,891) in \textbf{7,516.80 seconds}, as multiple SAT solver invocations for optimization increase runtime. PyEGo, limited to HG2.9K and SD due to its focus on single-file scripts, resolves 49\% of HG2.9K cases (1,435/2,891) in \textbf{1,817.40 seconds}. Its heuristic pre-selection of popular packages can prematurely eliminate valid dependency combinations, limiting applicability to multi-file projects and configuration files.

In the comparison with smartPip using a downgraded knowledge graph, SMTpip resolves more cases and does so more efficiently (Table~\ref{tab:downgraded_combined}). For example, on HG2.9K, SMTpip resolves 1,656/2,891 cases (58\%) in \textbf{263.36 seconds}, whereas smartPip resolves 1,640/2,891 cases (56\%) in \textbf{1,482.06 seconds}. A reason for smartPip’s lower success rate is that it intentionally narrows the set of candidate versions by internally rewriting some constraints: in particular, it replaces $\geq$ and $>$ constraints with the compatible-release operator $\sim=$, which permits only minor-version upgrades. For instance, suppose a project requires a dependency version constraint $\geq 1.4.5$, but smartPip internally treats it as $\sim= 1.4.5$. Under PEP~440~\cite{PEP440}, $\sim= 1.4.5$ restricts the selected version to the range $\geq 1.4.5$ and $< 1.5.0$, thereby excluding versions that still satisfy the original $\geq 1.4.5$ constraint. The original constraint does not impose this upper bound; thus, smartPip rejects valid versions $1.5.0$ and higher.

To better understand the unresolved cases, we manually analyzed SMTpip failures and categorized them into four groups. \emph{UNSAT Constraints} correspond to projects whose dependency and Python-version requirements are inherently contradictory, meaning that no valid environment exists. Resolving such cases would require modifying the project's requirements (e.g., upgrading or downgrading dependency constraints), which is beyond the scope of SMTpip and the compared techniques. SMTpip can report unsatisfiable dependency configurations to developers to help identify conflicting requirements; validating these conflicts directly with package developers is left for future work. \emph{Missing Metadata} occurs when the dependency knowledge graph lacks required dependency information for a package version due to incomplete metadata available from PyPI. \emph{Unsupported pip Features} account for approximately 5\% of unresolved cases and arise when projects rely on pip-specific functionality that is not yet supported by SMTpip, such as platform-specific dependency markers, direct URL dependencies, or local path dependencies. Finally, \emph{Solver Timeout} occurs when the SMT solver cannot complete within the imposed 500-second timeout because of extremely large dependency graphs and search spaces.

We also observed that SMTpip successfully resolves all cases that can be resolved by pip, Conda, smartPip, or PyEGo. Across all evaluated datasets, we did not encounter any instance where SMTpip failed while another tool succeeded. Even in benchmarks where SMTpip and pip resolve a similar number of projects, SMTpip demonstrates substantially better robustness on difficult dependency-resolution tasks. Pip's backtracking-based search strategy may struggle or time out when exploring very large dependency search spaces. For example, pip timed out on two projects in the WatchMan dataset and four projects in the Notebook dataset. One representative example is a project depending on \texttt{tableschema-py}, where SMTpip successfully resolved the dependency conflict and produced a valid environment in 22.41 seconds, whereas pip exceeded the 500-second timeout limit without finding a solution. These results demonstrate that SMTpip can efficiently handle challenging real-world dependency conflicts that are difficult for trial-and-error based dependency resolution approaches.

In addition to finding a compatible assignment when one exists, SMTpip explicitly reports when the constraint set is \emph{inconsistent} (i.e., no version selection can satisfy all declared dependency and interpreter constraints). This diagnostic is useful because it indicates that the installation failure is caused by mutually contradictory constraints that must be fixed upstream (e.g., by package maintainers or by correcting project requirements), rather than by trying additional candidate versions. In contrast, tools such as pip may expend substantial effort exploring candidates before failing, without clearly distinguishing an inherent inconsistency from search exhaustion; SMTpip returns an explicit UNSAT result for inconsistent instances.

\begin{table}[htbp]
\caption{Results of SMTpip, pip, Conda, and PyEGo in resolving dependency conflicts using latest dependency knowledge graphs.}
\label{tab:latest_combined}
\centering
\resizebox{7.5cm}{!}{
\begin{tabular}{|c|c|cc|}
\hline
\multirow{2}{*}{\textbf{Dataset}} &
  \multirow{2}{*}{\textbf{Tool}} &
  \multicolumn{2}{c|}{\textbf{Resolved Dependencies}} \\ \cline{3-4} 
 &
   &
  \multicolumn{1}{c|}{\textbf{Num}} &
  \makecell{\textbf{Total Time (sec)}} \\ \hline
\multirow{4}{*}{\textit{WatchMan}} &
  pip &
  \multicolumn{1}{c|}{112/159 (70\%)} &
  946.40 s \\ \cline{2-4} 
 &
  Conda &
  \multicolumn{1}{c|}{31/159 (19\%)} &
  352.22 s \\ \cline{2-4} 

 &
  \textbf{SMTpip} &
  \multicolumn{1}{c|}{\textbf{114/159 (72\%)}} &
  \textbf{226.86 s} \\ \hline
\multirow{5}{*}{\textit{HG2.9K}} &
  pip &
  \multicolumn{1}{c|}{1668/2891 (58\%)} &
  3185.20 s \\ \cline{2-4} 
 &
  Conda &
  \multicolumn{1}{l|}{1062/2891 (37\%)} &
  7516.80 s \\ \cline{2-4} 
 &
  PyEGo &
  \multicolumn{1}{c|}{1435/2891 (49\%)} &
  1817.40 s \\ \cline{2-4} 
 &
  \textbf{SMTpip} &
  \multicolumn{1}{c|}{\textbf{1668/2891 (58\%)}} &
  \textbf{433.68 s} \\ \hline
\multirow{5}{*}{\textit{SD}} &
  pip &
  \multicolumn{1}{c|}{76/100 (76\%)} &
  253.50 s \\ \cline{2-4} 
 &
  Conda &
  \multicolumn{1}{c|}{24/100 (24\%)} &
  482.8 s \\ \cline{2-4} 
 &
  PyEGo &
  \multicolumn{1}{c|}{62/100 (62\%)} &
84.52 s \\ \cline{2-4} 
 &
  \textbf{SMTpip} &
  \multicolumn{1}{c|}{\textbf{76/100 (76\%)}} &
  \textbf{36.48 s} \\ \hline
\multirow{4}{*}{\textit{Notebook}} &
  pip &
  \multicolumn{1}{c|}{887/1359 (65\%)} &
  5289.50 s \\ \cline{2-4} 
 &
  Conda &
  \multicolumn{1}{c|}{784/1359 (58\%)} &
  2414.90 s \\ \cline{2-4} 
 &

  \textbf{SMTpip} &
  \multicolumn{1}{c|}{\textbf{891/1359 (66\%)}} &
  \textbf{400.95 s} \\ \hline
\end{tabular}}
\end{table}

\begin{table}[htbp]
\caption{Results of SMTpip and smartPip in resolving dependency conflicts using downgraded dependency knowledge graphs.}
\label{tab:downgraded_combined}

\centering

\resizebox{7.5cm}{!}{
\begin{tabular}{|c|c|cc|}
\hline
\multirow{2}{*}{\textbf{Dataset}} &
  \multirow{2}{*}{\textbf{Tool}} &
  \multicolumn{2}{c|}{\textbf{Resolved Dependencies}} \\ \cline{3-4} 
 & 
  & \multicolumn{1}{c|}{\textbf{Num}} & \makecell{\textbf{Total Time (sec)}} \\ \hline
\multirow{2}{*}{\textit{WatchMan}} &
  smartPip &
  \multicolumn{1}{c|}{104/159 (65\%)} &
  533.50 s \\ \cline{2-4}
 &
  \textbf{SMTpip} &
  \multicolumn{1}{c|}{\textbf{110/159 (69\%)}} &
  \textbf{156.22 s} \\ \hline
\multirow{2}{*}{\textit{HG2.9K}} &
  smartPip &
  \multicolumn{1}{c|}{1640/2891 (56\%)} &
  1482.06 s \\ \cline{2-4}
 &
  \textbf{SMTpip} &
  \multicolumn{1}{c|}{\textbf{1656/2891 (58\%)}} &
  \textbf{263.36 s} \\ \hline
\multirow{2}{*}{\textit{SD}} &
  smartPip &
  \multicolumn{1}{c|}{62/100 (62\%)} &
  39.20 s \\ \cline{2-4}
 &
  \textbf{SMTpip} &
  \multicolumn{1}{c|}{\textbf{68/100 (68\%)}} &
  \textbf{24.42 s} \\ \hline
\multirow{2}{*}{\textit{Notebook}} &
  smartPip &
  \multicolumn{1}{c|}{689/1359 (50\%)} &
 393.03 s \\ \cline{2-4}&
  \textbf{SMTpip} &
  \multicolumn{1}{c|}{\textbf{711/1359 (53\%)}} &
  \textbf{305.73 s} \\ \hline
\end{tabular}
}
\end{table}

\begin{table*}[h]
\caption{Comprehensive time cost comparison across tools (pip, Conda, smartPip, PyEGo vs.\ SMTpip).
SC (Success Count) is the number of projects where both the tool and SMTpip successfully resolved dependency conflicts. TC (Time Cost) is the time taken for dependency resolution, shown as ``Tool TC | SMTpip TC'' with the speedup in parentheses (Tool TC / SMTpip TC). Speedup indicates how much faster SMTpip resolves dependencies compared to the respective tool.}
\label{tab:combined-comparison}
\resizebox{\textwidth}{!}{%
\begin{tabular}{|c|c|c|c|c|c|c|c|c|}
\hline
\multirow{2}{*}{\textbf{Dataset}} & \multicolumn{2}{c|}{\textbf{pip vs.\ SMTpip}} & \multicolumn{2}{c|}{\textbf{Conda vs.\ SMTpip}} & \multicolumn{2}{c|}{\textbf{smartPip vs.\ SMTpip}} & \multicolumn{2}{c|}{\textbf{PyEGo vs.\ SMTpip}} \\ \cline{2-9}
 & \textbf{SC} & \textbf{TC (pip | SMTpip)} & \textbf{SC} & \textbf{TC (Conda | SMTpip)} & \textbf{SC} & \textbf{TC (smartPip | SMTpip)} & \textbf{SC} & \textbf{TC (PyEGo | SMTpip)} \\ \hline
\textit{WatchMan} & 112 & 654.40 s | 226.86 s (\textbf{2.8$\times$}) & 31 & 138.22 s | 61.69 s (\textbf{2.2$\times$}) & 104 & 520 s | 151.84 s (\textbf{3.4$\times$}) & N/A & N/A \\ \hline
\textit{HG2.9K} & 1668 & 2165.20 s | 433.68 s (\textbf{4.9$\times$}) & 1062 & 4566.23 s | 276.12 s (\textbf{16.5$\times$}) & 1640 & 1476 s | 262.4 s (\textbf{5.6$\times$}) & 1435 & 1467.4 s | 346.84 s (\textbf{4.2$\times$}) \\ \hline
\textit{SD} & 76 & 198.50 s | 36.48 s (\textbf{5.4$\times$}) & 24 & 118.20 s | 11.52 s (\textbf{10.2$\times$}) & 42 & 25.2 s | 22.9 s (\textbf{1.1$\times$}) & 62 & 63.24 s | 29.76 s (\textbf{2.12$\times$}) \\ \hline
\textit{\begin{tabular}[c]{@{}c@{}} Notebook\end{tabular}} & 887 & 4568.50 s | 400.95 s (\textbf{11.3$\times$}) & 784 & 1936.50 s | 352.8 s (\textbf{5.4$\times$}) & 689 & 378.95 s | 296.27 s (\textbf{1.27$\times$}) & N/A & N/A \\ \hline
\textbf{Sum} & \textbf{2743} & \textbf{7586.6 s | 1097.97 s (\textbf{6.9$\times$})} & \textbf{1901} & \textbf{6759.15 s | 702.13 s (\textbf{9.6$\times$})} & \textbf{2475} & \textbf{2400.15 s | 733.45 s (\textbf{3.2$\times$})} & \textbf{1396} & \textbf{1530.64 s | 376.56 s (\textbf{4$\times$})} \\ \hline
\end{tabular}%
}
\end{table*}

\subsection{RQ2: Is the Technique Efficient Enough for Practical Use?}
\label{sec:rq2}

The efficiency of SMTpip in resolving package dependency conflicts is critical to its suitability for practical use in real-world software development. Thus, we analyze the speedup achieved by SMTpip compared to four established tools—pip, Conda, smartPip, and PyEGo—across four datasets: WatchMan, HG2.9K, SD, and Notebook. Table~\ref{tab:combined-comparison} presents a comprehensive time comparison for projects where both SMTpip and the compared tool successfully resolved dependency conflicts. Here, Success Count (SC) represents the number of projects where the dependency conflict was resolved by both tools, Time Cost (TC) denotes the total dependency-resolution time, and Speedup is computed as the ratio of the compared tool’s time cost to SMTpip’s.

Table~\ref{tab:combined-comparison} shows that SMTpip consistently achieves substantial speedups across all datasets. Overall, SMTpip is 6.9$\times$ faster than pip, 9.6$\times$ faster than Conda, 3.2$\times$ faster than smartPip, and 4$\times$ faster than PyEGo on the combined datasets. For example, on HG2.9K, SMTpip resolves 1,668 projects in 433.68 seconds, compared to pip’s 2,165.20 seconds (4.9$\times$ speedup). Pip’s longer times are due to iterative backtracking, which may explore many candidate versions before converging. Conda exhibits significant delays as well: on HG2.9K, SMTpip is 16.5$\times$ faster (4,566.23 seconds vs.\ 276.12 seconds for the 1,062 projects where both succeed), reflecting the additional overhead of Conda’s optimization criteria and repeated solver invocations.

In comparison with smartPip, SMTpip achieves a 5.6$\times$ speedup on HG2.9K (1,476 seconds vs.\ 262.4 seconds for the 1,640 projects where both succeed). SmartPip’s higher cost is due to the non-CNF encoding that increases solver overhead. PyEGo, evaluated only on HG2.9K and SD due to its focus on single-file Python scripts, shows SMTpip being 4.2$\times$ faster on HG2.9K (1,467.4 seconds vs.\ 346.84 seconds for the 1,334 projects where both succeed). In smaller datasets such as SD, SMTpip’s speedup over smartPip is modest (1.1$\times$), reflecting that simpler instances reduce the relative benefit of SMT-based global reasoning.

\subsection{
RQ3: Do the environments generated by SMTPip improve the executability of source code?}
\label{sec:rq3}

\begin{table}[t]
\centering
\caption{Execution validation results on HG2.9K and Notebook datasets with unified error categories.}
\label{tab:execution_combined}
\resizebox{\columnwidth}{!}{
\begin{tabular}{|l|c|c|c|c|c|c|}
\hline
\textbf{Dataset} &
\textbf{Tool} &
\textbf{Executed} &
\textbf{Success (\%)} &
\makecell{\textbf{Module/}\\\textbf{API Errors}} &
\makecell{\textbf{Interpreter}\\\textbf{Errors}} &
\makecell{\textbf{Other}\\\textbf{Errors}} \\ \hline

\multicolumn{7}{|c|}{\textbf{HG2.9K Dataset}} \\ \hline

HG2.9K & pip       & 1257 & 75.4\% & 287 & 102 & 22 \\ \hline
HG2.9K & smartPip  & 1235 & 75.3\% & 283 & 98  & 24 \\ \hline
HG2.9K & PyEGo     & 1331 & 92.7\% & 72  & 26  & 6  \\ \hline
HG2.9K & \textbf{SMTpip} & \textbf{1453} & \textbf{87.1\%} & \textbf{150} & \textbf{37} & \textbf{28} \\ \hline

\multicolumn{7}{|c|}{\textbf{Notebook Dataset}} \\ \hline

Notebook & pip       & 616 & 20.0\% & 920 & 345 & 1265 \\ \hline
Notebook & smartPip  & 554 & 18.0\% & 980 & 341 & 1205 \\ \hline
Notebook & PyEGo     & 770 & 25.0\% & 810 & 391 & 1110 \\ \hline
Notebook & \textbf{SMTpip} & \textbf{1230} & \textbf{39.92\%} & \textbf{592} & \textbf{280} & \textbf{979} \\ \hline

\end{tabular}
}
\end{table}

While RQ1 and RQ2 evaluate resolution success and speed, they do not assess whether the resolved virtual environments are \emph{functionally accurate}---that is, whether the selected dependencies actually allow the project to execute correctly.  Therefore, in RQ3 we evaluate the practical executability of the resolved environments. We performed execution evaluation on both the HG2.9K and Notebook datasets because their entry points can be identified reliably. The WatchMan and SD datasets were excluded from execution testing because they consist of multi-file projects, making it difficult to reliably determine the appropriate entry point for automated execution.

For each program gist in the HG2.9K dataset whose environment was successfully resolved by a tool, we created an isolated virtual environment, installed the selected package versions, and executed the program entry point. Execution outcomes were categorized as: (1) successful execution; (2) \emph{Module/API failures}, typically caused by deprecated packages or API-breaking changes; (3) \emph{Interpreter failures}, including installation errors due to incompatible Python versions or \texttt{SyntaxError} from unsupported language features; and (4) \emph{Other} failures, including OS-specific dependencies, external services, tool-specific integrations, missing packages, or absent datasets and configuration files.

Table~\ref{tab:execution_combined} summarizes the results. Although pip and SMTpip resolve the same number of environments (1668), SMTpip yields substantially more successful executions (1453 vs.\ 1257). This improvement is primarily due to interpreter selection. Because most HG2.9K gists specify dependencies only through \textit{requirements.txt}, Python version requirements are typically absent (Figure~\ref{fig:config_files}). As a result, pip resolves dependencies using the currently installed interpreter without searching for alternative Python versions. In contrast, SMTpip incorporates \texttt{Requires-Python} metadata and jointly resolves interpreter and package version constraints, enabling the selection of a compatible interpreter. Consequently, SMTpip achieves an execution success rate of 87.1\%, compared to 75.4\% for pip and 75.3\% for smartPip. Although PyEGo attains a higher success rate (92.7\%), it resolves fewer environments (1435 vs.\ 1668), resulting in fewer successful executions overall (1331 vs.\ 1453).

SMTpip also substantially reduces interpreter-related failures (37 vs.\ 102 for pip and 98 for smartPip). This reduction is largely due to interpreter handling: when projects are specified only via \textit{requirements.txt}, pip and smartPip resolve dependencies under the already-installed interpreter and do not compute or recommend an alternative Python version; as a result, interpreter incompatibilities may appear as installation failures or \texttt{SyntaxError} caused by unsupported language features. Across all tools, most remaining failures stem from module/API mismatches caused by breaking changes in major-version upgrades, while others arise from host-specific libraries, proprietary infrastructure, or missing external artifacts that fall outside the scope of dependency resolution.

We also evaluated SMTpip on the Notebook dataset. Table~\ref{tab:execution_combined} summarizes the results. The dataset comprises 1,359 real-world Jupyter Notebook projects with dependency conflicts and contains 3,081 \texttt{.ipynb} files, as each project typically includes multiple notebooks. We executed all 3,081 notebooks using environments resolved by all tools. SMTpip achieves the highest execution success rate of 39.92\% (1,230/3,081), outperforming pip (20.0\%), smartPip (18.0\%), and PyEGo (25.0\%). The most common failure categories for SMTpip were \texttt{Module/API errors} (592 cases), \texttt{Interpreter errors} (280 cases), and \texttt{Other errors} (979 cases), where the latter includes file-not-found issues, type/syntax errors, and missing external resources such as datasets and configuration files. Similar failure patterns are observed for baseline tools, although SMTpip consistently reduces interpreter-related failures due to improved interpreter selection and constraint handling.

\section{Discussion}
\label{sec:discussion}
This section discusses questions related to our study.

\noindent\textbf{Number of Packages in the Install Scripts:}
SMTpip applies an optimization function that prunes unnecessary packages from generated install scripts. This pruning step removes packages that are not required to satisfy the project's dependency constraints, producing a more compact install script. The summary statistics of the package counts produced by SMTpip and smartPip in four datasets are reported in Table~\ref{tab:package_counts}. As shown in Table~\ref{tab:package_counts}, smartPip consistently produces larger install scripts across all four datasets. Some projects use only Python’s standard library (no third-party packages), but we keep them because selecting a compatible Python interpreter can still matter. To illustrate this difference with a concrete example, consider a simple requirements.txt file that specifies only \textit{click ==} 6.6 and \textit{pip-tools $\geq$} 4.0.0. After dependency resolution, the solver selects \textit{click ==} 6.6 and \textit{pip-tools ==} 4.2.0, where \textit{pip-tools ==} 4.2.0 declares only two runtime dependencies (\textit{click $\geq$} 6 and \textit{six}). Consequently, SMTpip produces a minimal install script containing exactly three packages: \textit{click} 6.6, \textit{pip-tools} 4.2.0, and \textit{six} 1.16.0. In contrast, more recent releases of \textit{pip-tools} such as version 7.4.1 declare a substantially larger set of dependencies including \textit{build $\geq$} 1.0.0, \textit{click $\geq$} 8, \textit{pip $\geq$} 22.2, \textit{pyproject-hooks}, \textit{setuptools}, and \textit{wheel}. Although smartPip resolves to the same version \textit{pip-tools} 4.2.0, it expands the install script by including 20 packages, most of which are unnecessary.\\

\begin{table}[ht]
\centering
\caption{Summary statistics of package counts in install scripts produced by SMTpip and smartPip across datasets. ``Additional Packages'' = smartPip count minus SMTpip count.}
\label{tab:package_counts}
\resizebox{9cm}{!}{%
\fbox{%
\begin{tabular}{l|c|c|c}

\textbf{Dataset} & \textbf{SMTpip} & \textbf{smartPip} & \textbf{Additional Packages} \\
\midrule
& Min/Max/Median/Avg & Min/Max/Median/Avg & Min/Max/Median/Avg \\
\midrule
Watchman   & 0/53/3/5.52     & 0/107/6/11.72    & 0/54/3/6.20   \\
HG2.9K     & 0/33/1/2.50     & 0/76/2/5.66      & 0/49/1/3.15   \\
Notebook   & 0/93/1/3.71     & 0/151/3/8.31     & 0/74/2/4.61   \\
SD         & 0/59/6/11.09    & 0/144/16/24.55   & 0/90/9/13.46  \\

\end{tabular}%
}}
\end{table}

\noindent\textbf{Effect of Dependency Graph Size on Performance:}
We examined whether dependency graph size affects resolution time using the Notebook dataset, which contains the largest number of projects. The dataset was divided into four quartiles (Q1--Q4) based on dependency graph size, and for each group we measured the time required to generate SMT expressions and to solve them. Table~\ref{tab:graph_size_metrics} summarizes the quartile boundaries and timing statistics.

Generation time increases gradually from smaller graphs (Q1--Q2) to medium graphs (Q3), and then rises sharply for the largest graphs (Q4): the median generation time increases from 0.01~s in Q2 to 0.07~s in Q3 and reaches 2.82~s for Q4 projects. Solving time follows a similar but steeper pattern, with the largest graphs producing the most pronounced increases.

Spearman rank-order correlations confirm these trends. Generation time correlates strongly with dependency graph size, \(r_s[1359] = .697,\, p < .001\), while solving time shows an even stronger association, \(r_s[1359] = .901,\, p < .001\). These results indicate that larger dependency graphs consistently lead to longer generation and solving times, with solver time exhibiting the dominant growth for larger graphs.
\\

\begin{table}
  \centering
  \scriptsize
  \caption{Dependency-graph-size groups and timing statistics (seconds).}
  \label{tab:graph_size_metrics}
  \begin{tabular}{l| r| r}
    \toprule
    Quartile & \makecell{ Generating the SMT expression \\ (mean/median/min/max)} 
               & \makecell{ Solving the SMT expression \\ (mean/median/min/max)}
  \\
    \midrule
    Q1   & 0.02 / 0.01 / 0.00 / 0.11 & 0.00 / 0.00 / 0.00 / 0.02  \\
    Q2   & 0.02 / 0.01 / 0.00 / 0.12 & 0.01 / 0.01 / 0.00 / 0.03 \\
    Q3   & 0.03 / 0.02 / 0.00 / 0.14 & 0.17 / 0.07 / 0.03 / 0.57  \\
    Q4   & 0.13 / 0.10 / 0.03 / 0.75 & 3.50 / 2.82 / 0.45 / 10.01  \\
    \bottomrule
  \end{tabular}
\end{table}

\noindent\textbf{Rationale for Direct CNF Encoding:}
SMTpip generates SMT constraints directly in conjunctive normal form (CNF), rather than constructing higher-level non-CNF formulas and relying on the solver to perform conversion. This design choice is motivated by both theoretical considerations from SAT/SMT solving and empirical observations. In particular, CNF is expected to perform well because modern SAT/SMT solvers deal with CNF internally. We also evaluated several non-CNF variants and observed consistently worse performance in practice; therefore, we present the CNF encoding used by SMTpip as our primary design. SMTpip uses a compact Boolean encoding with one variable per package--version, enforcing single-version semantics via cardinality constraints and encoding dependencies as implications. Our empirical comparisons against non-CNF variants support this rationale, showing that the CNF encoding used by SMTpip consistently performs best in practice.

\section{Threats to Validity}\label{THREATS}
This section discusses threats to the validity of our research. 

\noindent\textbf{External Validity:} Threats to external validity refer to the generalizability of our findings. We evaluate SMTpip using projects written in Python. One can argue that the results may not be generalized to other Python projects. However, we would like to point to the fact that we consider four different datasets of varying sizes and covering different application domains. While Watchman, HG2.9K and SD datasets were used by prior studies, we create a new dataset consisting of real-world Jupyter Notebook projects collected from GitHub based on several criteria. The core of our technique does not depend on any specific programming language or ecosystem. Thus, the technique should be applicable to other ecosystems with minor changes.\\

\noindent\textbf{Internal Validity:} Threats to internal validity relate to potential biases or inaccuracies in our methodology and measurements. First, SMTpip relies on the correctness and completeness of package metadata, which may be incomplete or inconsistent in practice. Second, entry-point detection and runtime execution may be affected by environment-specific behavior and nondeterminism (e.g., timeouts or network-dependent code). Third, execution-based evaluation does not guarantee functional correctness, since HG2.9K and Notebook datasets lack ground-truth outputs or test cases, making it impossible to verify program correctness. Finally, results may vary with different solver configurations; we use default Z3 settings for consistency, and all comparisons are performed under identical experimental conditions.

\section{Conclusion}
\label{conclusion}

Dependency conflicts and interpreter incompatibilities in modern Python development hinder reproducibility and slow environment setup. We presented \textbf{SMTpip}, an SMT-based framework that models environment setup as a global version-selection problem under both package and interpreter compatibility constraints. SMTpip uses a compact encoding with a weighted Max-SMT formulation to efficiently compute constraint-consistent and executable environments. We evaluated SMTpip on four real-world datasets, including 1,359 GitHub Jupyter Notebook projects and three established benchmarks. SMTpip achieves substantial speedups---6.9$\times$ over pip, 9.6$\times$ over Conda, 3.2$\times$ over smartPip, and 4$\times$ over PyEGo---while maintaining strong resolution coverage. Execution-based evaluation further shows that SMTpip produces environments with higher executability and significantly fewer interpreter-related failures compared to baseline tools. Overall, these results demonstrate that interpreter-aware, constraint-based global reasoning can improve both the efficiency and executability of Python dependency resolution in practice. Future work includes extending the framework to ecosystems with different installation semantics and exploring incremental resolution for evolving environments.

\bibliographystyle{IEEEtran}
\bibliography{references}

\end{document}